\documentclass[aps,prx,preprintnumbers,amssymb,amsfonts,floatfix,letterpaper,twoside,twocolumn]{revtex4}

\usepackage{natbib}
\usepackage{graphicx}
\usepackage[utf8]{inputenc}
\usepackage{mathptmx}
\usepackage{epsfig,amsopn}
\usepackage{graphicx}
\usepackage{amsmath,amssymb}
\usepackage{natbib,color}
\usepackage{braket}
\usepackage{float}
\usepackage{enumerate}
\usepackage[normalem]{ulem}
\usepackage{todonotes}
\allowdisplaybreaks[1]

\newcommand{\eref}[1]{Eq.~(\ref{#1})}%
\newcommand{\fref}[1]{Fig.~\ref{#1}} %

\def\bea{\begin{eqnarray}}
\def\eea{\end{eqnarray}}

\begin{document}

\title{Search with home returns provides advantage under high uncertainty}

\author{Arnab Pal$^{1,2}$}
\email{arnabpal@mail.tau.ac.il}

\author{\L{}ukasz Ku\'smierz$^{3}$}

\author{Shlomi Reuveni$^{1,2}$}
\email{shlomire@tauex.tau.ac.il}

\affiliation{\noindent \textit{$^{1}$School of Chemistry, The Center for Physics and Chemistry of Living Systems, \& The Mark Ratner Institute for Single Molecule Chemistry, Tel Aviv University, Tel Aviv 6997801, Israel}}

\affiliation{\noindent \textit{$^{2}$The Raymond and Beverly Sackler Center for Computational Molecular and Materials Science, Tel Aviv University, Tel Aviv 6997801, Israel}}

\affiliation{\noindent \textit{$^{3}$Laboratory for Neural Computation and Adaptation, RIKEN Center for Brain Science, 2-1 Hirosawa, Wako, Saitama 351-0198, Japan}}

\date{\today}

\begin{abstract}
\noindent

Many search processes are conducted in the vicinity of a favored location, i.e., a home, which is visited repeatedly. Foraging animals return to their dens and nests to rest, scouts return to their bases to resupply, and drones return to their docking stations to recharge or refuel. Yet, despite its prevalence, very little is known about search with home returns as its analysis is much more challenging than that of unconstrained, free-range, search. Here, we develop a theoretical framework for search with home returns. This makes no assumptions on the underlying search process and is furthermore suited to treat generic return and home-stay strategies. We show that the solution to the home-return problem can then be given in terms of the solution to the corresponding free-range problem---which not only reduces overall complexity but also gives rise to a simple, and universal, phase-diagram for search. The latter reveals that search with home returns outperforms free-range search in conditions of high uncertainty. Thus, when living gets rough, a home will not only provide warmth and shelter but also allow one to locate food and other resources quickly and more efficiently than in its absence.

\end{abstract}

\maketitle

\noindent

\section{Introduction}

Consider a falcon roaming the sky in search of prey well hidden amongst the grass below. The falcon will wander around for a while, but if prey is not found it will eventually return to its nest empty-handed. Other animals---humans included---display similar behaviour while foraging and when engaged in search activities  \cite{HR1,HR2,HR3}; and home-return capabilities are now routinely built into robots and drones to avoid running out of fuel or battery power. However, while the observation that most natural search processes are home-bound goes back to Darwin \cite{Darwin}, it is still unclear if this situation merely reflects the prevalence of permanent dwellings, or rather is a result of evolutionary convergence to a superior search strategy. To start answering this question, one must first understand how being home-bound affects search and the time it takes to locate a target. In what follows, we analyze this problem and characterize precisely under which circumstances having a home allows one to locate food and other resources quickly and more efficiently than in its absence.

\begin{figure*}[t!]
\begin{center}
\includegraphics[width=17.5cm]{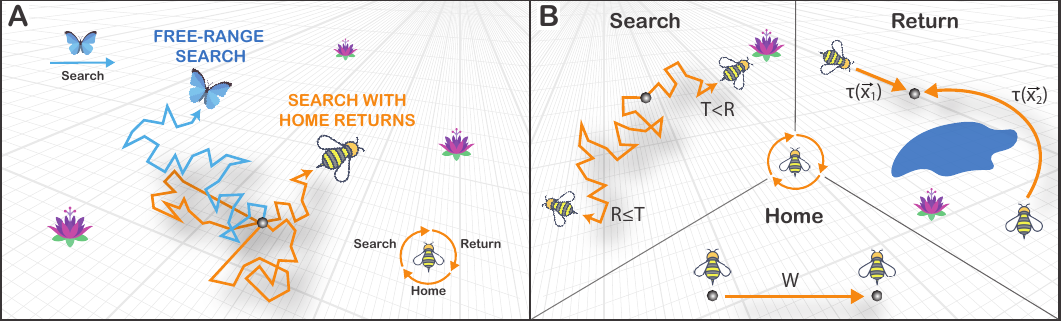}
\caption{A. Free-range search vs. search with home returns. The butterfly (free-range searcher) and the bee (searcher with home returns) set off in search of a flower (target). The butterfly, which has no permanent dwellings, will look for a flower until it finds one. In contrast, if the bee is unable to find a flower it will return to its hive, spend some time there, and start searching again at a later occasion. B. Search with home returns is a cyclic, three stage, process. In the search stage a target is sought for a time that is the minimum of the free-range FPT, $T$, and the restart time $R$. If $T<R$, a target is found and the search ends. Conversely, if $R \leq T$, no target is found and the searcher heads back home. The duration of the return stage, $\tau(\vec{x})$, is determined by the searcher's position $\vec{x}$ at the end of the search phase. This stage ends when the searcher is back home. In the home stage the searcher stays home for a time $W$.}
\label{slow-resetting-main}
\end{center}
\end{figure*}

A free-range searcher will set off from a certain location and look for a target until it is found. In contrast, search with home returns is a cyclic process which consists of three stages: search, return, and home (\fref{slow-resetting-main}A). How much time does it take such a searcher to find its target? At face value, it seems that this question can be answered by taking advantage of the existing theory of search \cite{Search1,Search2,Search3,Search4,Search5,Search6,Search7,Search8} and first-passage \cite{FPT1,FPT2,FPT3,FPT4,FPT5,FPT6,FPT7,FPT8}, and of recent advancements in our understanding of first-passage under restart  \cite{Restart-ND-0,Restart-ND-1,Restart-ND-2,Restart-ND-3,Restart-ND-4,Restart-ND-5,Restart-ND-6,Restart-ND-7,Restart-ND-8,Restart-ND-9,Restart-ND-10,Restart-D-0,Restart-D-1,Restart-D-2,Restart-D-3,Restart-D-4,Restart-D-5,Restart-D-6,Restart-D-7,Restart-D-8}. Indeed, search with home returns can be seen as a regular first-passage process that is restarted by home returns. However, basic models of first-passage under restart are a far cry from reality as they assume that home-returns are instantaneous and that home-stays can also be neglected \cite{Restart-ND-0,Restart-ND-1,Restart-ND-2,Restart-ND-3,Restart-ND-4,Restart-ND-5,Restart-ND-6,Restart-ND-7,Restart-ND-8,Restart-ND-9,Restart-ND-10}. 

More sophisticated models of search with home returns lump together return and home-stay times assuming that the search stage is followed by some generic delay \cite{Restart-D-0,Restart-D-1,Restart-D-2,Restart-D-3,Restart-D-4,Restart-D-5,Restart-D-6,Restart-D-7}. This is a step in the right direction: it takes time to get from one place to another, and time spent home to e.g., recover, recharge, or refuel, may not be negligible. However, the time it takes a searcher to return home will typically depend on the distance home, as places that are further away take more time to be reached. Yet, this basic physics is clearly ignored when assuming that the delay which follows the search stage is generic and independent of the searcher's position as it starts heading back home \cite{return1,return2,return3,return4,return5}. This non-realistic modelling assumption is in many ways similar to the complete decoupling between waiting time and jump length in the continuous time random walk (CTRW) model \cite{CTRW1,CTRW2,CTRW3,CTRW4}. In the latter case, the problem was solved by the development of space-time coupled CTRWs \cite{SPC-CTRW} and L\'evy walks \cite{Walks-CTRW1,Walks-CTRW2,Walks-CTRW3,Walks-CTRW4,Walks-CTRW5,Walks-CTRW6} which introduced explicit correlations between time and distance traveled. In what follows, we take a similar approach and build a space-time coupled theory for first-passage under restart. This, in turn, will be used to describe search with home returns.

The paper is structured as follows. In Secs. \ref{TF} and \ref{TFE}, we develop a  theoretical framework for search with home returns. We show that this framework provides a practical analysis tool as it allows one to provide the solution to the home return problem in terms of the solution to the corresponding problem without home returns. This useful property is demonstrated in Sec. \ref{DS} with the example of diffusive search. In Sec. \ref{PD}, we build on our framework to reveal a universal phase diagram for search. In particular, we show that search with home returns is preferable in conditions of high uncertainty as it can then reduce the mean time taken to locate a target. This property is illustrated with the example of drift-diffusive search in Sec. \ref{DDSM}. In Sec. \ref{opt}, we show that search with home returns can also reduce fluctuations in the time taken to locate a target. This feature is illustrated with the example of L\'evy search in Sec. \ref{levy} and its importance is discussed. Conclusions and outlook are given in Sec. \ref{conc}. Some details and derivations are relegated to the appendix.

\section{A theoretical framework for search with home returns}\label{TF} Consider a searcher that starts at the origin (home) of a (possibly infinite) $d$-dimensional arena at time zero. In the absence of home returns, the searcher will locate one of the existing targets in the arena following a random time $T$. This time is a property of the free-range problem, and we will henceforth refer to it as the free-range first-passage time (FPT). We will not make any assumptions on the arena, the search process, and target distribution that govern $T$. However, and in contrast to free-range search, here we will consider a situation where the searcher returns home if it fails to locate the target within a time $R$ (can be random) which we will henceforth refer to as the restart time. Thus, if $T<R$ the searcher finds the target before it is required to return and the search process completes. Otherwise, the searcher will stop looking for the target and start its return back home (\fref{slow-resetting-main}B - Search).

The time it takes the searcher to return home will typically depend on the searcher's position at the end of the search stage (\fref{slow-resetting-main}B - Return). For example, the searcher may return home by moving at a constant speed along the shortest possible path. The return time is then simply given by the distance to home over the speed of travel. However, various constraints, e.g., topographic ones, may force the searcher to follow a different route and may also affect its velocity. Such situations will result in more complicated relations between the position of the searcher and its return time. To capture this, we allow the return time $\tau(\vec{x})$ to be a general function of the searcher's position $\vec{x}$. After the searcher returns home it stays there for some generic time $W$ which can also be random (\fref{slow-resetting-main}B - Home). This, search--return--home, cycle repeats itself until a target is found at some point during the search stage. In what follows, we will assume that targets cannot be located during the return and home phases (see discussion in Sec. \ref{conc}). 

The above description allows us to write a renewal equation for the FPT of search with home returns, which is the time it takes the searcher to locate a target. Denoting this time by $T_R$, we have
\begin{equation}
\begin{array}{l}
T_{R}=\left\{ \begin{array}{lll}
T &  & \text{if ~~}T<R\text{ ,}\\
 & \text{ \ \ }\\
R+\tau(\vec{x})+W+T_R'&  & \text{if~~ }R\leq T\text{ ,}
\end{array}\right.\text{ }\end{array}
\label{renewal-1-main}
\end{equation}
where $T$, $R$, $\tau(\vec{x})$, and $W$ were defined above; and $T_R'$ is an independent and identically distributed copy of $T_R$. Taking expectations in \eref{renewal-1-main}, we obtain (Appendix \ref{APP_Eq2})
\begin{align}
\langle T_R  \rangle= \underbrace{  \frac{\langle \text{min}(T,R) \rangle}{\text{Pr}(T<R)}}_{\text{search}}+\underbrace{\frac{\langle I(R \leq T)\tau(\vec{x})  \rangle}{\text{Pr}(T<R)}}_{\text{return}}+
\underbrace{\frac{\text{Pr}(R \leq T)\langle W \rangle}{\text{Pr}(T<R)}}_{\text{home}},
\label{MFPT-1}
\end{align}
where $I(R \leq T)$ is an indicator function which takes the value one if $R \leq T$, i.e., with probability $\text{Pr}(R \leq T)$, and is zero otherwise; and different contributions to the sum are labeled according to their source.  

The first term on the right-hand side of \eref{MFPT-1} gives the FPT of the searcher in an idealized scenario where return and home times can be neglected ($\tau(\vec{x})=0$, $W=0$) \cite{Restart-ND-5}. The second term gets its contribution from the time it takes the searcher to return home and the third term comes from the time spent at home. Evaluating the first and third terms is straightforward given the probability distributions of $R,~T$, and $W$ (Appendix \ref{APP_EvalEq2}). The second term is slightly more delicate because it depends on $\vec{x}$---the random position of the searcher at the end of the search stage. To evaluate this term, we let $f_R(t)$ denote the probability density function of the restart time $R$. We then observe that
\begin{align}
\langle I(R \leq T)\tau(\vec{x})  \rangle &= \int_0^\infty dt f_R(t)~ \langle \tau(\vec{x}(t)) I(R \leq T)|R=t \rangle \nonumber \\
&=  \int_0^\infty dt f_R(t)\text{Pr}(T \geq t) \langle \tau(\vec{x}(t)) \big|R=t,T \geq t \rangle,
\label{indicator}
\end{align}
where we have first conditioned on restart happening at time $t$, and then on $T$ being either smaller or larger than this time. Note that a non-zero contribution is obtained only for $T\geq t$, i.e., only when the target is not found and a return actually takes place. 

\begin{figure*}[t!]
\begin{center}
\includegraphics[width=17.5cm]{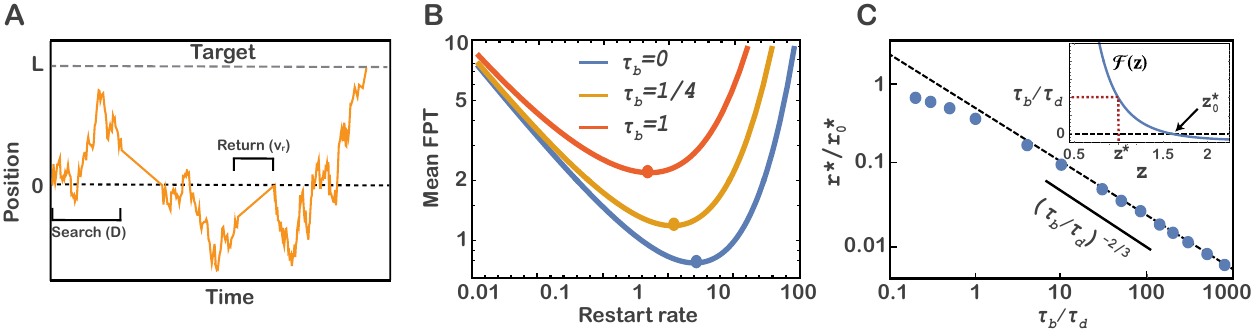}
\caption{A. An illustration of diffusive search with home returns. B. The mean FPT, $\langle T_r \rangle$, from \eref{MFPT-main} vs. the restart rate $r$. Here, $\tau_d=1/2$ and results are shown for different values of $\tau_b$ [see \eref{time-scales}]. C. The scaled optimal restart rate, $r^*/r_0^*$, obtained from a minimization of \eref{MFPT-main} vs. $\tau_b/\tau_d$. The scaling predicted by \eref{optimalr*scaling} is seen to hold. Inset. $\mathcal{F}(z)$ from \eref{optimal-1} vs. $z$. The ratio $\tau_b/\tau_d$ sets the solution $z^*$.}
\label{diffusive home-range}
\end{center}
\end{figure*}

In order to proceed, we define the free-range propagator, $G_0(\vec{x},t)$, as the probability density to find the searcher at position $\vec{x}$ at time $t$ given that it started at the origin. Note that this propagator is called free-range because it is defined in the presence of targets but in the absence of home-returns. Thus, the free-range survival probability is given by $\text{Pr}(T\geq t)=\int_{\mathcal{D}}d\vec{x}~G_0(\vec{x},t)$, where $\mathcal{D}$ is the available search domain. The internal expectation in \eref{indicator} can then be written as $\langle \tau(\vec{x}(t)) \big|R=t,~T \geq t \rangle=\frac{1}{ \text{Pr}(T \geq t)}\int_{\mathcal{D}}~d\vec{x}~\tau(\vec{x})~G_0(\vec{x},t)$. Substituting this expression into \eref{indicator}, we obtain 
\bea
\langle I(R \leq T)\tau(\vec{x})  \rangle= \int_0^\infty dt~f_R(t)
\int_{\mathcal{D}} d\vec{x}~\tau(\vec{x})~G_0(\vec{x},t)~.
\label{second-expectation}
\eea
Equation (\ref{second-expectation}) asserts that the second term in \eref{MFPT-1} can be evaluated given the free-range propagator $G_0(\vec{x},t)$, which in turn allows full evaluation of the mean FPT. 

Starting from \eref{renewal-1-main}, and proceeding similarly to the above, the distribution of the FPT, $T_R$, can also be determined. Letting  $\tilde{T}_R(s)=\langle e^{-sT_R}\rangle$ stand for the  Laplace transform of the latter, we find (Appendix \ref{APP_General_Case}) 
\bea
\tilde{T}_R(s)=\frac{\text{Pr}(T<R)\tilde{T}_{\text{min}}(s)}{1- \tilde{W}(s)  \int_0^{\infty}~dt~f_{R}(t)e^{-st}~\int_{\mathcal{D}}~ d\vec{x}~e^{-s\tau(\vec{x})}~G_0(\vec{x},t)}~\nonumber, \\
\label{LT-genericR_main}
\eea
with $\tilde{W}(s)=\langle e^{-sW} \rangle$ standing for the Laplace transform of $W$, and $\tilde{T}_{\text{min}}(s)=\langle e^{-sT_{\text{min}}} \rangle$ standing for the Laplace transform of the random variable $T_{\text{min}}=\{ T|T<R \}$ whose density is given by $f_{T_{\text{min}}}(t) = \frac{f_T(t) \int_t^\infty~dt'~f_R(t')}{\text{Pr}(T<R)}=\frac{f_T(t) \text{Pr}(R>t)}{\text{Pr}(T<R)}~$. Equation (\ref{LT-genericR_main}) asserts that the distribution of $T_R$ can be determined given the free-range propagator $G_0(\vec{x},t)$, and the random variables $R$ and $W$. In addition, all the moments can be computed using the formula $\langle T_R^n \rangle=(-1)^n\frac{d^n}{ds^n} \tilde{T}_R(s)|_{s=0}$.

\section{Exponential restart times}\label{TFE}

So far, we have made no assumptions on the distribution of the time $R$ which governs restart. In what follows, we show that much insight can be gained by focusing on the case where $R$ is exponentially distributed with rate $r$. Letting $\tilde{G}_0(\vec{x},r)=\int_0^\infty~ dt~e^{-rt}~G_0(\vec{x},t)$ and $\tilde{T}(r)=\int_0^\infty~ dt~e^{-rt}~f_T(t)=1-r\int_{\mathcal{D}}d\vec{x}~\tilde{G}_0(\vec{x},r)$ stand respectively for the Laplace transforms of $G_0(\vec{x},t)$ and $f_T(t)$ evaluated at $r$, we find that in this case \eref{MFPT-1} boils down to (Appendix \ref{APP_Eq5})
\bea
\langle T_r  \rangle 
=\underbrace{ \frac{1-\tilde{T}(r)}{r \tilde{T}(r)}}_{\text{search}}+\underbrace{\frac{1-\tilde{T}(r)}{ \tilde{T}(r)} \langle  \tau(\vec{x}) \rangle_r}_{\text{return}}+\underbrace{ \frac{1-\tilde{T}(r)}{ \tilde{T}(r)} \langle W \rangle}_{\text{home}} ,
\label{MFPT-2}
\eea
where $\langle \tau(\vec{x})\rangle_r \equiv \int_{\mathcal{D}}~d\vec{x}~\tau(\vec{x})~\phi_r(\vec{x})$ is the mean return time taken with respect to the probability measure $\phi_r(\vec{x})=\tilde{G}_0(\vec{x},r)/\int_{\mathcal{D}}d\vec{x}~\tilde{G}_0(\vec{x},r)$. Similarly, the expression for the FPT distribution can also be simplified to read (Appendix \ref{APP_Eq6})
\bea
\tilde{T}_r(s)=\frac{\tilde{T}(s+r)}{1-r~\tilde{W}(s) ~\int_{\mathcal{D}}~d \vec{x}~e^{-s\tau(\vec{x})}~ \tilde{G}_0(\vec{x},s+r) }.
\label{TR-Laplace-transform}
\eea 
From \eref{TR-Laplace-transform} we see that the distribution of $T_r$ can always be written in terms of the Laplace transforms of the free-range propagator $G_0(\vec{x},t)$, and the random variables $T$ and $W$. 

\section{Diffusive search with home returns}\label{DS}
To illustrate how the framework developed above can be utilized in practice, we examine a paradigmatic case study. Consider a 1-d search process in which a particle that starts at the origin diffuses until it hits a stationary target; and let $D$ and $L$ denote respectively the diffusion constant and the initial distance from the target. In addition, assume that the process is restarted at a constant rate $r$ upon which the searcher returns home at a constant speed $v_r$ (\fref{diffusive home-range}A). In what follows, the time spent home will be neglected as its stand-alone contribution is already well-understood \cite{Restart-D-0,Restart-D-1,Restart-D-2,Restart-D-3,Restart-D-4,Restart-D-5,Restart-D-6,Restart-D-7}.

To progress, we recall that
the free-range propagator of this  problem is given by \cite{FPT1}
\bea
G_0(x,t)=\frac{1}{\sqrt{4\pi D t}} \left( e^{ -\frac{x^2}{4Dt} }-e^{ -\frac{(2L-x)^2}{4Dt} } \right)~.
\label{G0-main}
\eea
To get the mean FPT with home returns, we observe that the time penalty due to a ballistic home-return from position $x$ is given by $\tau(x)=|x|/v_r$. Plugging in the above into \eref{MFPT-2} gives (Appendix \ref{APP_Eq8})
\bea
\langle T_r  \rangle =\underbrace{\frac{1}{r}\left( e^{\sqrt{\tau_d r}} -1 \right)}_{\text{search}}+\underbrace{ \tau_b~\left[ \frac{2 \sinh(\sqrt{\tau_d r})}{\sqrt{\tau_d r}} -1 \right]}_{\text{return}}~,
\label{MFPT-main}
\eea
where
\bea
\tau_d=\frac{L^2}{D}~\text{,~~~and}~~~\tau_b=\frac{L}{v_r}~,
\label{time-scales}
\eea
stand respectively for the diffusive and ballistic time scales in the problem. 

In the limit $\tau_b \to 0$, \eref{MFPT-main} boils down to the classical result for the mean FPT of diffusion with  resetting \cite{Restart-ND-1}, but we would now like to understand the effect of non instantaneous and space-time-coupled home returns. In \fref{diffusive home-range}B, we plot $\langle T_r \rangle$ as a function of the restart rate for $\tau_d=1/2$ and different values of $\tau_b$ (see Appendix \ref{APP_Eq8_cor} for numerical corroboration of these results). We then observe that diffusive search with home returns is always superior to diffusive free-range search---regardless of how slow home returns are. This can also be seen directly from \eref{MFPT-main} by noting that $\langle T_r \rangle$ there is finite for $r>0$, but diverges for $r=0$ where the searcher  does not return home.

Diving deeper, we observe that two things happen as we increase the ballistic (return) time scale: (i) it takes more time for the searcher to locate the target, i.e., $\langle T_r\rangle$ becomes larger; and (ii) the optimal restart rate, $r^*$, which minimizes $\langle T_r\rangle$ becomes smaller. The first effect is easy to understand by inspection of the return term in \eref{MFPT-main}. Quantitative analysis of the second effect reveals a non trivial scaling relation. 

When $\tau_b = 0$, the optimal restart rate $r_0^*$ can be determined by minimizing the first term in \eref{MFPT-main}. One then finds \cite{Restart-ND-1}: $r_0^*=z^{*2}_0/\tau_d$ with $z^*_0=1.593...$ standing for the solution to the following  transcendental equation
$1-e^{-z}-\frac{z}{2}=0$.
Minimizing $\langle T_r  \rangle$ in \eref{MFPT-main} for $\tau_b>0$, we find that this result generalizes to give $r^*={z^*}^2/\tau_d$ with $z^*$ standing for the solution to transcendental equation (Appendix \ref{APP_Eq10})
\bea
\mathcal{F}(z)\equiv \frac{2}{z^2}\frac{1-e^{-z}-\frac{z}{2}}{(1-\frac{1}{z})+(1+\frac{1}{z})e^{-2z}}=\frac{\tau_b}{\tau_d}.
\label{optimal-1}
\eea
Noting that $z^*$ is uniquely determined by the ratio $\tau_b/\tau_d$ on the right hand side of \eref{optimal-1} (\fref{diffusive home-range}C, inset), we conclude that  $r^*/r_0^*={z^*}^2/z^{*2}_0$.

In the limit $\tau_b \ll \tau_d$, one has $r^*/r_0^* \approx 1$ by definition. In the other extreme $\tau_b \gg \tau_d$, which in turn implies $z^*\to 0$ (\fref{diffusive home-range}C inset). Expanding $\mathcal{F}(z)$ around $z=0$, we find $\mathcal{F}(z) = \frac{3}{2 z^3} + O(\frac{1}{z})$ (Appendix \ref{APP_Exp}). Equating this with $\tau_b/\tau_d$ on the right side of \eref{optimal-1} we conclude that (\fref{diffusive home-range}C)
\begin{equation}
\begin{array}{l}
r^*/r_0^* \simeq\left\{ \begin{array}{lll}
1 &  & \text{for ~~}\tau_b \ll \tau_d \text{ }\\
 & \text{ \ \ }\\
\left(\frac{3}{2z_0^{*3}} \right)^{2/3}~\left(\frac{\tau_b}{\tau_d}\right)^{-2/3}&  & \text{for~~ }\tau_b \gg \tau_d\text{ .}
\end{array}\right.\text{ }\end{array}
\label{optimalr*scaling}
\end{equation}
We thus see that the interplay between search and home-returns gives rise to a power law which governs the optimal restart rate for 1-d diffusive search with home returns. Consequently, by substituting  \eref{optimalr*scaling} into \eref{MFPT-main}, we find that the optimal mean FPT obeys (Appendix \ref{APP_Eq_12})
\begin{equation}
\begin{array}{l}
\langle T_{r^*}  \rangle  \sim \left\{ \begin{array}{lll}
\tau_d &  & \text{for ~~}\tau_b \ll \tau_d \text{ }\\
 & \text{ \ \ }\\
\tau_b &  & \text{for~~ }\tau_b \gg \tau_d\text{ .}
\end{array}\right.\text{ }\end{array}
\label{optimalTr*scaling}
\end{equation}
And so, while arbitrary restart rates may easily lead to a situation where $\langle T_{r}  \rangle \gg \text{max}(\tau_b,\tau_d)$, the optimal mean FPT  asymptotically scales like $\langle T_{r^*}\rangle \sim \text{max}(\tau_b,\tau_d)$.

\begin{figure*}[t!]
\includegraphics[width=17.5cm]{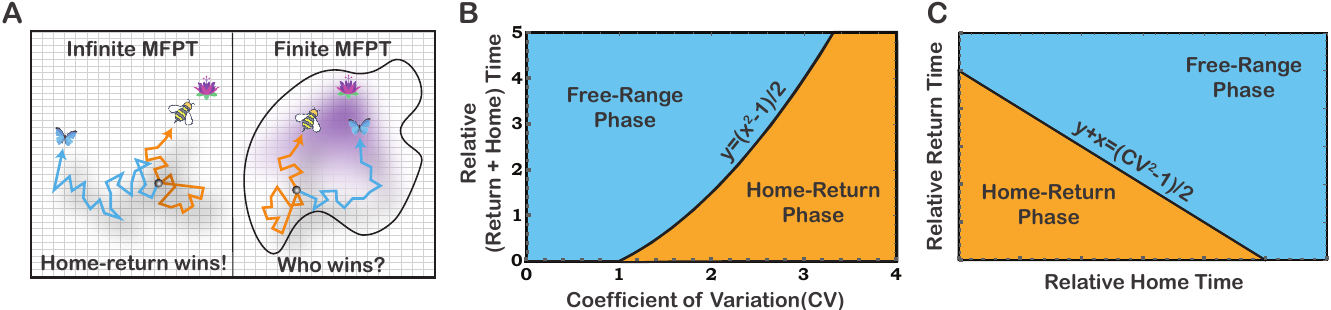}
\caption{A. Search with home returns wins over free-range search whenever the mean FPT of the latter diverges. However, when the free-range mean FPT is finite, e.g., when the search arena is finite or when motion is biased in the direction of the target, either free-range search or search with home returns can have a lower mean FPT. B \& C. The phase-space determined by \eref{criterion-main} is spanned by three dimensionless parameters: the coefficient of variation of the free-range FPT, $CV=\sigma(T)/\langle T \rangle$, the relative mean return time, $\langle \tau(\vec{x})\rangle_0/\langle T \rangle$, and the relative mean home time $\langle W \rangle/\langle T \rangle$. When system parameters belong to the home-return (free-range) phase the introduction of home-returns is asserted to decrease (increase) the mean FPT to the target.}
\label{home-range-free-range}
\end{figure*}

\section{A phase-diagram for search}\label{PD}
The above example illustrates a situation where search with home returns offers significant performance advantage over free-range search. To generalize, one only needs to observe that since the mean FPT in \eref{MFPT-2} is finite for $r>0$ (under mild regularity conditions: $\langle W \rangle<\infty~,\int_{\mathcal{D}} d\vec{x}~\tau(\vec{x})~G_0(\vec{x},t)<\infty$)---search with home returns offers a huge performance advantage in all conditions where the mean FPT of the underlying free-range process diverges (\fref{home-range-free-range}A left). This suggests that search with home returns performs best when search conditions are at their worst, but how to quantify and further extend this statement to situations where the underlying free-range FPT has a finite mean is not immediately clear as either free-range search or search with home returns may perform better (\fref{home-range-free-range}A right).

When does the introduction of home returns to a free-range search process lower the mean FPT to the target? To answer this question, one should take  $\langle T_r \rangle$ in \eref{MFPT-2} and check when  $d\langle T_r \rangle/dr|_{r = 0}<0$, which we find happens when (Appendix \ref{derivation13}) 
\bea
CV^2 > 1+\frac{2 \langle \tau(\vec{x})\rangle_0}{\langle T  \rangle} + \frac{2\langle W \rangle }{\langle T \rangle}~.
\label{criterion-main}
\eea
Here, $\langle T \rangle$ and $CV=\sigma(T)/\langle T \rangle$ are the mean and relative standard deviation (coefficient of variation) of the  free-range FPT, $\langle W \rangle$ is the mean home-stay time, and $\langle \tau(\vec{x})\rangle_0 = \int_{\mathcal{D}}~d\vec{x}~\tau(\vec{x})~\phi_0(\vec{x})=\frac{1}{\langle T  \rangle} \int_{\mathcal{D}}~d\vec{x}~\tau(\vec{x})~\tilde{G}_0(\vec{x},0)$ is the mean return time in the limit $r\to0$.

The condition in Eq. (\ref{criterion-main}) relates three dimensionless quantities and reveals that search with home returns outperforms free-range search in conditions of high uncertainty. Indeed, on the left hand side of Eq. (\ref{criterion-main}) stands the $CV$ which quantifies the relative magnitude of fluctuations, or uncertainty, around the free-range mean FPT. These fluctuations need to be large in order for the introduction of home-returns to be beneficial. On the right hand side of the inequality stand the relative mean return time, $\langle \tau(\vec{x})\rangle/\langle T \rangle$, and the relative mean home time, $\langle W \rangle/\langle T \rangle$, which act as penalties against home returns and set the bar for the critical magnitude of fluctuations at which the transition between the free-range phase and home-return phase occurs. The resulting phase-diagram for search is graphically illustrated in panels B \& C of \fref{home-range-free-range}. 

\section{Drift-Diffusive search with home returns}\label{DDSM}

To demonstrate how the universal result in Eq. (\ref{criterion-main}) manifests itself in a concrete example, we consider a simple model for search in the presence of guidance cues. Namely, we consider the same diffusive search with home returns as in \fref{diffusive home-range}A above, but now assume that the particle also drifts at an average velocity $v$. Note that when the particle drifts away from the target ($v<0$) the free-range mean FPT diverges and search with home returns is always preferable (see discussion above). We thus focus on the $v>0$ case which could e.g., model search in the presence of an attractant (potential field) that biases the searcher's motion in the direction of the target.

\begin{figure}[t]
    \centering
\includegraphics[height=3.25cm,width=8.5cm]{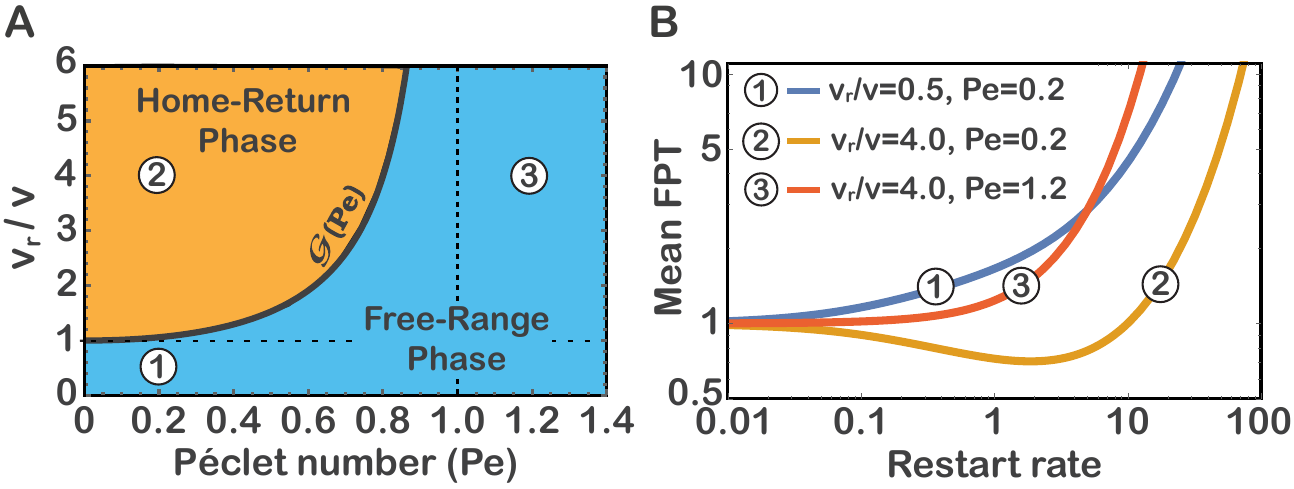}
\caption{A. The phase space of drift-diffusive search as determined by \eref{critical}. The free-range phase and home-return phase are separated by $\mathcal{G}(Pe)$ from \eref{Pe_main}. B. The mean FPT of drift-diffusive search with home returns vs. the restart rate (see Appendix \ref{DDS_Cor} for details and corroboration via numerical simulations). Here, $L=v=1$, and other  parameters are set by position in phase-space (numbered circles). When system parameters belong to the home return phase, e.g., for curve number (2), the introduction of home-returns decreases the mean FPT to the target. The converse happens for curves (1) \& (3) whose parameters belong to the free-range phase.}
\label{critical-phase-plot}
\end{figure}

The free-range propagator of drift-diffusion in the presence of an absorbing boundary  (target) is known to be given by \cite{FPT1}
\bea
G_0(x,t)=\frac{1}{\sqrt{4\pi D t}}\left[ e^{-\frac{(x-vt)^2}{4Dt}}-e^{\frac{L v}{D}}e^{-\frac{(x-2L-vt)^2}{4Dt}} \right].
\label{propagator-drift}
\eea
To build the search phase space, we first write all terms in \eref{criterion-main} in terms of the natural parameters of drift-diffusion. Setting off from \eref{propagator-drift}, a straightforward calculation gives $\langle T \rangle=L/v$ and $CV^2=1/Pe$, where $Pe=Lv/2D$ is the P\'eclet number \cite{FPT1}, i.e., the ratio between the rates of advective and diffusive transport. In addition, we find $\langle \tau(\vec{x}) \rangle_0=\frac{L}{2v_r}\left( 1-e^{-2Pe}-Pe+Pe^2 \right)/Pe^2$, with $v_r$ standing once again for the home-return speed (Appendix \ref{DDS}). 

When $Pe\geq1$ drift rules over diffusion which means that guidance cues towards the target are strong. Uncertainty in the free-range FPT is then relatively small and the condition in \eref{criterion-main} cannot be satisfied since $CV^2=1/Pe\leq1$. On the other hand, when $0<Pe<1$, diffusion rules over drift which means that guidance cues towards the target are weak. Uncertainty in the free-range FPT is then larger and we find that the condition in \eref{criterion-main} is satisfied whenever (Appendix \ref{DDS}, \fref{critical-phase-plot})  
\bea
v_r>v \cdot \mathcal{G}(Pe)~,
\label{critical}
\eea
with 
\bea
\mathcal{G}(Pe)=\frac{1-e^{-2Pe}}{Pe(1-Pe)}-1.
\label{Pe_main}
\eea
This means that the introduction of home returns will be beneficial whenever the return speed $v_r$ is greater than a critical speed $v_r^*=v \cdot \mathcal{G}(Pe)$.  Measured in units of the drift velocity $v$, the critical return speed is uniquely determined by the P\'eclet number and hence by the relative uncertainty in the free-range FPT. When $Pe\ll1$, $v_r^*\approx v$, but in the limit $Pe \to1$, we have $v_r^* \sim v/(1-Pe)$. Thus, as guidance cues (drift) towards the target become stronger the return speed must increase sharply in order for search with home returns to remain beneficial.

\section{Optimal search with home returns reduces mean and variance of time to target}\label{opt}

When fluctuations in the free-range FPT are high such that the inequality in \eref{criterion-main} holds, the introduction of home returns is asserted to lower the mean FPT to the target. This, in turn, implies the existence of an optimal restart rate $r^*>0$ for which the mean FPT, $\langle T_{r^*}  \rangle$, is minimal. Reduction of the mean FPT is clearly important, but large fluctuations around the mean FPT can be deleterious as living organisms rely on a steady supply of nutrients and other essential resources. To this end, we now show that optimal search with home returns provides another important advantage: it reduces the variance of the FPT to the target.

Fluctuations around $\langle T_{r^*}  \rangle$ have contributions coming from all stages of search, but note that those coming from the home stage are exclusively controlled by the searcher and can thus be made small. In fact, it is enough to require that $\sigma (W) \leq \langle W \rangle$ to show that the condition in \eref{criterion-main} implies  
\bea
\sigma(T_{r^*})^2 \leq \langle T_{r^*} \rangle^2+ 2\langle T_{r^*} \rangle[\langle \tau(\vec{x})\rangle_{*} + \langle W \rangle] ~,
\label{conc_1}
\eea
where the mean return time $\langle \tau(\vec{x})\rangle_*$ is computed like $\langle \tau(\vec{x})\rangle_{0}$ in \eref{criterion-main}, but with respect to the measure $\phi_{*}(\vec{x})=\tilde{G}_{*}(\vec{x},0)/\langle T_{r^*}  \rangle$ such that
\bea
\langle \tau(\vec{x})\rangle_*=\frac{1}{\langle T_{r^*} \rangle}\int_{\mathcal{D}}d\vec{x}~\tau(x)\tilde{G}_*(\vec{x},0),
\eea
where $\tilde{G}_*(\vec{x},0)=\int_0^\infty~dt~G_*(\vec{x},t)$ and $G_{*}(\vec{x},t)$ is the propagator of the search process with home returns conducted at the optimal restart rate $r^*$.

Equation (\ref{conc_1}) is proven by contradiction. Assume this equation does not hold, and observe that this implies $\frac{\sigma(T_{r^*})^2}{\langle T_{r^*} \rangle^2} > 1+\frac{2 \langle \tau(\vec{x})\rangle_*}{\langle T_{r^*} \rangle} + \frac{2\langle W \rangle }{\langle T_{r^*} \rangle}~.$ Now, the condition in \eref{criterion-main} asserts that the mean FPT time $\langle T_{r^*} \rangle$ can be lowered by restarting the entire search process at a small rate $\epsilon$. However, since the search stage is already being restarted at a rate $r^*$, the introduction of an additional restart rate $\epsilon$ amounts to restarting this stage at a rate $r^*+\epsilon$. Contrary to the search stage, the return and home stages are not restarted at a rate $r^*$. Thus, one only needs to  consider what happens when both these stages are restarted at a rate $\epsilon$.

If the searcher is in the return stage it must have gotten there due to a restart event. Assuming that this restart event caught the searcher at some position $\vec{x}$, it will take the searcher $\tau(\vec{x})$ units of time to return home. In this return, the searcher will take a path that connects $\vec{x}$ with the origin (home). Consider a point $\vec{y}$ along this path, and let $\tau_{\vec{x}}(\vec{y})$ denote the remaining return time of a searcher which passes through $\vec{y}$ in his way back home from  $\vec{x}$. In general, $\tau_{\vec{x}}(\vec{y})$ need not be equal to $\tau(\vec{y})$, i.e., to the time it takes the searcher to return from $\vec{y}$ when restart happens there. However, demanding that return times and paths obey $\tau_{\vec{x}}(\vec{y})=\tau(\vec{y})$ for every starting point $\vec{x}$ and every point $\vec{y}$ along a return path is very natural. Indeed, this only means that the time it takes the searcher to get back home from $\vec{y}$ does not depend on how it got there in the first place; and note that when this is the case restarting the return phase has no effect on the overall dynamics. Specifically, if a restart event catches the searcher during the return phase at a point $\vec{y}$ along a path connecting $\vec{x}$ with the origin, the searcher will take $\tau(\vec{y})$ units of time to return home from $\vec{y}$ which is exactly what would have happened in the absence of a restart event. We thus conclude that restarting the return phase at a rate $\epsilon$ has no effect, i.e., it is equivalent to not restarting the return phase at all.

Now, consider what happens when restart occurs during the home phase. Since $\tau(\vec{x}=\vec{0})=0$ by definition, the return time has zero contribution and one then only needs to understand the direct effect restart has on the duration of the time spent home. Recall that the home phase is on average $\langle W \rangle$ units of time long. Thus, if a restart event occurs during the home phase it will, on average, force the searcher to spend an additional  $\langle W \rangle$ units of time at home. This time should be compared to the time the searcher would have spent home if restart would not have occurred at the moment it did. This time is known as the residual life time of $W$ \cite{Gallager}, and renewal theory teaches us that its mean is given by $\langle W_{\text{res}} \rangle=\frac{\langle W^2 \rangle}{2\langle W \rangle}=\frac{\sigma^2(W) +\langle W \rangle^2}{2\langle W \rangle}$. For example, if $W$ is deterministic, i.e., has zero variance, $\langle W_{\text{res}} \rangle= \langle W \rangle/2$ as restart would on average ``catch'' the searcher half way through its home stay duration. More generally, $\langle W_{\text{res}} \rangle$ can be smaller or larger than $\langle W \rangle$, but note that when $\sigma(W)<\langle W \rangle$ we always have $\langle W_{\text{res}} \rangle< \langle W \rangle$. Thus, when the standard deviation of the home stay time is smaller than its mean, restart will (on average) tend to prolong home stays as it ``replaces'' $\langle W_{\text{res}} \rangle$ with $\langle W \rangle$ which is longer.

From the above we conclude that if the addition of a small restart rate $\epsilon$ to all stages of search lowers the mean FPT below $\langle T_{r^*} \rangle$, then  the addition of a small restart rate $\epsilon$ to the search phase only will also lower the mean FPT. Indeed, restarting the return phase is equivalent to not restarting it at all. In addition, not restarting the home phase (instead of restarting it at a rate $\epsilon$) will result in shorter home stays (provided that $\sigma(W)<\langle W \rangle$), which will lower the mean FPT even more as the target cannot be found while sitting at home. We thus find that $\langle T_{r^*+\epsilon} \rangle<\langle T_{r^*} \rangle$ which is, however, in contradiction to optimality as $r^*$ is defined to be the restart rate that brings $\langle T_{r} \rangle$ to a minimum. Concluding, we see that assuming that \eref{conc_1} does not hold leads to a contradiction, which means that this equation must hold.

From the above we draw an important conclusion. While there is no fundamental upper limit on fluctuations of free-range FPTs, those of optimal search with home returns must obey the bound in \eref{conc_1}. Moreover, since  $\langle T_{r^*} \rangle < \langle T  \rangle$ by definition of the optimal restart rate $r^*$, we conclude that the combination of Eqs. (\ref{criterion-main}) and (\ref{conc_1}) gives  
\bea
\sigma(T_{r^*}) < \sigma(T) ~,
\label{conc_2}
\eea
whenever $\langle \tau(\vec{x})\rangle_* \leq \langle \tau(\vec{x})\rangle_0$. The latter condition is expected to hold in the generic case since a searcher that returns home from time to time will typically be found closer to home than one that does not. Thus, in addition to lowering the mean FPT to the target, optimal search with home returns also leads to a net reduction of fluctuations around the mean. 

\section{L\'evy search with home returns}\label{levy}

To illustrate the double advantage conferred by search with home returns, we consider a L\'evy walker that conducts search in a finite two-dimensional arena with multiple targets. L\'evy walks \cite{Walks-CTRW1,Walks-CTRW2,Walks-CTRW3,Walks-CTRW4,Walks-CTRW5} have been widely applied to model animal foraging and motion \cite{HR3,Search1,Search2,Search3,Search4,Search6,Search7} as there are cases where they provide advantage over diffusive search strategies \cite{Search2,Search6,Search7,Search8,Walks-CTRW2,Levy-opt1}. It has thus been hypothesized that natural selection favours L\'evy walks, which may explain their prevalence in nature. In what follows we show that the L\'evy search strategy can be further improved when it is combined with home-returns. We start with a brief review of L\'evy walks.

In the basic version of the L\'evy walk model, a random walker travels along a straight line at a constant speed for some random time. At the end of the excursion, the walker randomly chooses a new direction of motion and travels along it (at the same speed) for another random duration before it turns  again. The model is thus characterized by the travel speed $v_{LW}$ and the distribution of the random times between turning points. The latter are taken to be independent and identically distributed, and further assuming a finite mean and variance leads to motion that is asymptotically diffusive. However, when considering L\'evy walks one is usually interested in cases where the long time asymptotics of the travel time probability density has a power-law form $\psi(\tau) \sim  \tau^{-1-\alpha}$ with $0<\alpha<2$. This form leads to a diverging second moment and superdiffusive motion.

\begin{figure}[t]
    \centering
\includegraphics[width=8.7cm]{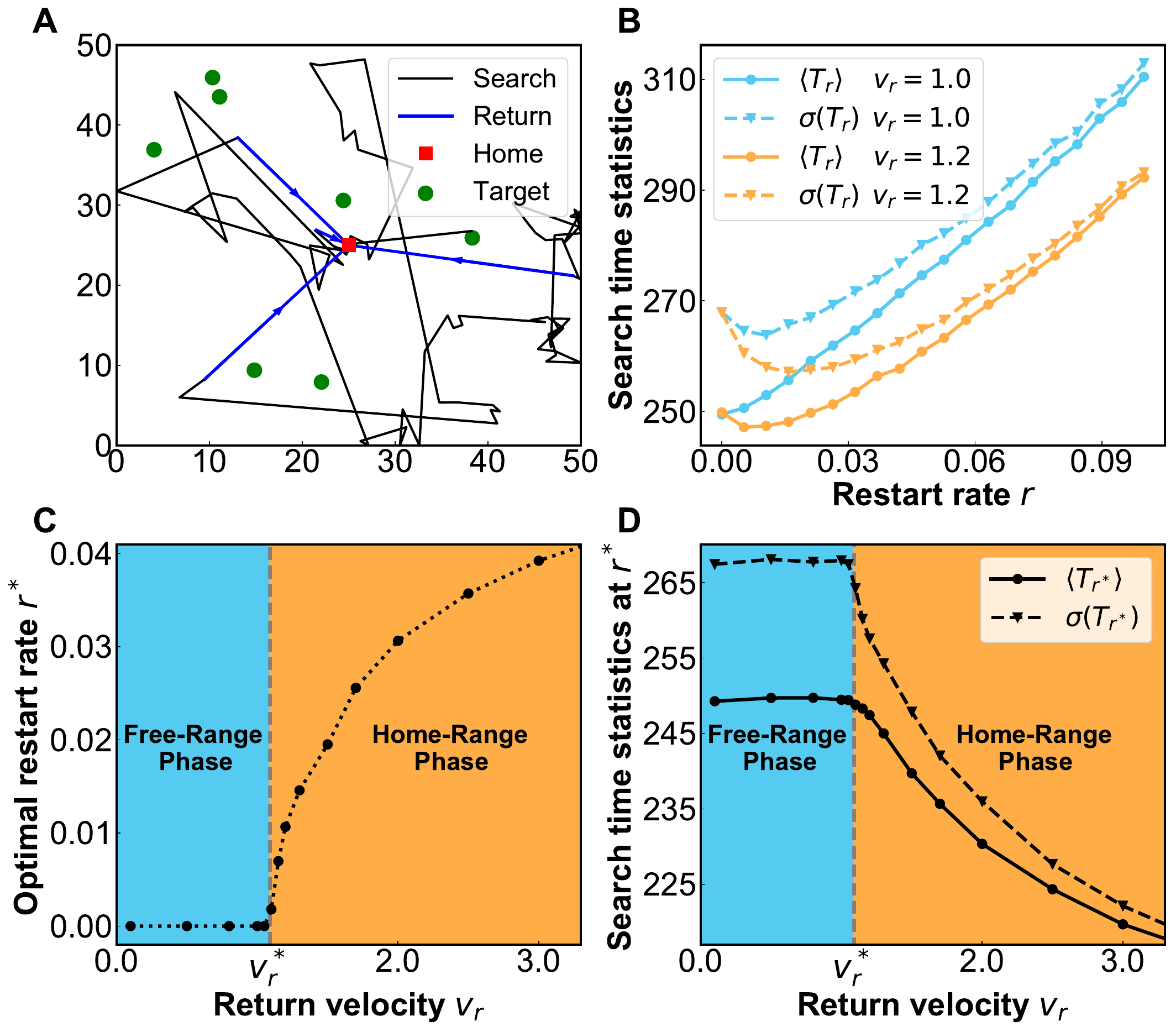}
\caption{Optimal search with home returns reduces the mean and variance of the time to target. A. Here, this general feature is demonstrated for an agent foraging within a bounded two-dimensional arena. The agent performs a truncated L\'evy walk with steps taken from a heavy-tailed distribution $\rho(l)\propto l^{-2}$. The search process is restarted at a rate $r$, and home returns are conducted at a constant speed $v_r$. The process ends when any one of the targets is found. B. The mean (circles) and standard deviation (triangles) of the first passage time vs. the restart rate for two different return speeds. C. The optimal restart rate for which the mean FPT $\langle T_{r}\rangle$ is minimized vs. the return speed. The critical return speed above which $r^*>0$ is estimated via \eref{critical_est}. This method gives $v_r^* \simeq 1.09$ (dashed vertical line), which is in excellent agreement with independent numerical simulations of the home-return process for different values of $v_r$ (circles). D. The mean (circles) and standard deviation (triangles) of the first passage time under optimal restart vs. $v_r$. For $v_r>v_r^*$, both the mean and standard deviation of the FPT are strictly lower than those found for free-range search.}
\label{double-win}
\end{figure}

In Fig. \ref{double-win}A, we consider an agent whose task is to locate any one of seven targets that were placed randomly in a square arena. The agent conducts L\'evy search with home returns. In the search phase, the agent performs a L\'evy walk with $v_{LW}=1$. After each step the direction and length of the following step are chosen at random. The direction is drawn from the uniform angle distribution, whereas the step length is given by $l=v_{LW} \tau$, where $\tau=\tau_0/Z$ represents the random duration till the next turn. Taking $Z$ to be uniformly distributed on the unit interval $(0,1]$, one can show that $\psi(\tau) = \tau_0/\tau^{2}$ ($\tau>\tau_0$) which gives $\alpha=1$ \cite{Walks-CTRW6}. The probability density function governing the step length is then given by $\rho(l) = v_{LW} \tau_0/l^2$ for $l> v_{LW} \tau_0$, and we take $\tau_0=1$. Finally, in order to account for the finite size of the arena, we note that if the L\'evy walker arrives at a boundary its step is truncated and a new step is generated. Thus, in practice, we consider a truncated L\'evy walk \cite{Walks-CTRW2}.

The L\'evy walk described above is restarted at a rate $r$, and home-returns are performed along the shortest possible path with a constant return speed $v_r$. To map the phase space of this search process, we scanned multiple $(v_r, r)$ pairs. For each pair, we simulated $N=10^6$ sample trajectories that end when any one of the targets is hit during the search stage (recall that we assume that targets cannot be found during the return stage). In Fig. \ref{double-win}B, we plot the mean and standard deviation of the resulting FPT vs. the restart rate for two different values of the return speed $v_r=1$ and $v_r=1.2$. When $r=0$, search is conducted in the absence of home returns and we find  that $\sigma(T)>\langle T \rangle$. Thus, in this example $CV>1$ and \eref{criterion-main} asserts that the mean FPT can be lowered by the introduction of home-returns; provided the return speed is high enough (here we take $W=0$). Indeed, for $v_r=1.2$ we see that the mean FPT is minimized at $r^*>0$. However, the optimal restart rate for $v_r=1$ is $r^*=0$, which suggests that the critical return speed (above which home returns become beneficial) is somewhere in the range $1<v_r^*<1.2$.

The critical return speed $v_r^*$ can be determined via numerical evaluation of the mean return time $\langle \tau(\vec{x})\rangle_0$ in \eref{criterion-main}. To do this, we only need to simulate the search process without home returns ($r=0$). For each linear segment of the  L\'evy walk we note two quantities: the segment's duration $\tau_i$, and the average distance of that segment to the home position (starting point), which we denote as $d_i$. Although such an average distance between a line segment and a point can in principle be calculated analytically, here we estimate its value by averaging over $10$ regularly spaced points along the segment (faster numerically). We then calculate the average return distance as 
\begin{equation} 
\overline{d} = \sum\limits_{i} d_i \tau_i/\sum\limits_i \tau_i ~.
\end{equation}
The mean return time in \eref{criterion-main} is then given by $\langle \tau(\vec{x}) \rangle_0=\overline{d}/v_r$, and substituting back into \eref{criterion-main} gives
$CV^2 > 1+\frac{2 }{\langle T  \rangle}\frac{\overline{d}}{v_r}.$
Rearranging, we find that the critical return speed is given by 
\begin{equation}
    v_r^* = \frac{2 \overline{d}}{\sigma(T)^2/\langle T \rangle - \langle T \rangle},\label{critical_est}
\end{equation}
which is uniquely determined by the free-range search process.

In Fig. \ref{double-win}C, we compare the estimate obtained from \eref{critical_est} to an independent estimate of $v_r^*$. The latter is obtained by direct numerical evaluation of the optimal restart rate. For a given return speed $v_r$, the optimal restart rate is found in two steps: (1) among the values of $r$ that were simulated we find the one that gives the shortest mean FPT, and then (2) we fit a quadratic function to the estimated mean FPT as a function of $r$ for six to eleven data points (adaptive algorithm) around the value found in (1). The quadratic function is used to predict the values of $r^*$ and $\langle T_{r^*} \rangle$. If the value of $r^*$ is predicted to be smaller than or equal to zero we take $r^*=0$ and the corresponding mean FPT as the optimal. The critical return velocity can then be determined as the smallest return velocity for which $r^*>0$. As can be seen, this method of estimating $v_r^*$ is in excellent agreement with the prediction coming from \eref{critical_est}. Finally, for the optimal restart rates found, we plot the mean and standard deviation of the FPT vs. the return speed (Fig. \ref{double-win}D). For $v_r>v_r^*$, the mean and standard deviation are found to be lower than the values obtained in the absence of home-returns. Thus, optimal search with home returns reduces both the mean and variance of the time to target as predicted in the previous sections.

\section{Conclusions and outlook}\label{conc}

Search with home returns is widely observed in nature, but its analysis has so far been challenging. We developed a theoretical framework for this process and used it to show that solutions to first-passage problems with home-returns can always be given in terms of solutions to the  corresponding free-range first-passage problems, i.e., those without home returns. The latter are known for a plethora of cases as first-passage time problems have been studied for decades; but even when this is not the case, the framework developed herein is still useful as it reduces a complicated problem to a much simpler one. Most importantly, our framework reveals a simple, and universal, phase-diagram for search. This, in turn, can be used to decide under which circumstances search with home returns is preferable to free-range search. 

Our framework advances the field of first-passage under restart in several directions. First and foremost, it allows for a realistic description of restart  by accounting for non-instantaneous and space-time coupled returns. To this end, the searcher's return time was allowed to be an arbitrary function of its position at the restart moment, which naturally couples returns to the underlying stochastic motion. The latter can be general, which is also true for the distributions of restart and home-waiting times. Thus, our framework is applicable to a large variety of stochastic search processes, in arbitrary dimensions, and generally shaped domains that contain either single or multiple targets. Specifically, we provided general results for the mean [Eqs. (\ref{MFPT-1}) and (\ref{MFPT-2})] and distribution [Eqs. (\ref{LT-genericR_main}) and (\ref{TR-Laplace-transform})] of the first-passage time of a search process with home returns, and further demonstrated how these results apply to several case studies of interest.

To further elucidate the effect of home-returns, we asked under which conditions adopting this strategy is advantageous to search. We showed that this question can be answered based on the statistical properties of the underlying first-passage process, i.e., that which is conducted without home returns. This, in turn, gave us a phase diagram for search and revealed that search with home returns outperforms free-range search in conditions of high uncertainty. Specifically, the introduction of home returns will lower the mean FPT to a target whenever the relative magnitude of the fluctuations, or uncertainty, around the free-range mean FPT is large [\eref{criterion-main}]. Moreover, under the same conditions, optimal search with home returns will also reduce the fluctuations around the mean FPT [\eref{conc_2}], which is important as living organisms rely heavily on a steady supply of nutrients and other essential resources. Indeed, even when the time taken to locate a resource is, on average, short enough to support life---large fluctuations around the average are deleterious and may result in death. Thus, search with home returns offers a double advantage, which unequivocally asserts the superiority of this strategy when facing uncertainty conditions.

While the prevalence of search with home returns in organisms ranging from insects to humans is probably due to the amalgamation of many contributing factors, our analysis shows that having a home may also be important as it allows one to locate food and other resources quickly and more efficiently than in its absence. Importantly, we find that this is true even when knowledge on the surrounding environment is not taken into account, and despite the fact that our analysis assumed that targets cannot be located in the return stage, i.e., while returning home. Thus, in reality, search with home returns is expected to perform even better than predicted here. Free-range search may out-compete search with home returns, but only in conditions of low uncertainty. This suggests that search with home returns may have evolved as a bet-hedging strategy that performs best when search conditions are at their worst.

\subsection*{Acknowledgments}
The authors would like to acknowledge Tam\'{a}s Kiss, Sergey Denisov and Eli Barkai, organizers of the 672. WE-Heraeus Seminar: ``Search and Problem Solving by Random Walks'', as discussions that led to this work began there.  Shlomi Reuveni would like to deeply acknowledge Sidney Redner for a series of joint discussions which led to this work. The authors would also like thank Guy Cohen and Ofek Lauber for commenting on early versions of this manuscript. Shlomi Reuveni acknowledges support from the Azrieli Foundation, from the Raymond and Beverly Sackler Center for Computational Molecular and Materials Science at Tel Aviv University, and from the Israel Science Foundation (grant No. 394/19). Arnab Pal acknowledges support from the Raymond and Beverly Sackler Post-Doctoral Scholarship at Tel-Aviv University.
 
\appendix 

\section{Detailed derivation of \eref{MFPT-1} in the main text}\label{APP_Eq2}
\noindent
To derive \eref{MFPT-1} in the main text, we first rewrite \eref{renewal-1-main} to obtain
\bea
T_{R}=\text{min}(T,R)+I(R \leq T) \left[ \tau(\vec{x})+W+T_R'  \right]~,
\eea
where $I(R \leq T)$ is an indicator function that takes the value one if $R \leq T$, and zero otherwise.
Taking expectations on the both sides of the above equation, 
we obtain
\bea
\langle T_R  \rangle=\langle \text{min}(T,R) \rangle+ \langle I(R \leq T) \left[ \tau(\vec{x})+W+T_R'  \right]  \rangle.
\eea
Recalling that $W$ and $T_R'$ are independent of $T$ and $R$, and that  $\langle I(R \leq T)\rangle=\text{Pr}(R\leq T)$ by definition, we find
\begin{align}
\langle T_R  \rangle&=\langle \text{min}(T,R) \rangle+\langle I(R \leq T)\tau(\vec{x})  \rangle+
\text{Pr}(R\leq T) \langle W \rangle \nonumber \\
&+\text{Pr}(R\leq T) \langle T_R' \rangle.
\end{align}
Finally, as $T_R'$ is an independent and identically distributed (IID) copy of $T_R$ we have $\langle T_R  \rangle=\langle T_R'  \rangle$, and simple rearrangement then gives
\begin{align}
\langle T_R  \rangle=\frac{\langle \text{min}(T,R) \rangle}{\text{Pr}(T<R)}+\frac{\langle I(R \leq T)\tau(\vec{x})  \rangle}{\text{Pr}(T<R)}+\frac{\text{Pr}(R\leq T) \langle W \rangle}{\text{Pr}(T<R)}.
\label{SMFPT-2}
\end{align}
Equation (\ref{SMFPT-2}) is equivalent to \eref{MFPT-1} in the main text. Substituting \eref{second-expectation} into \eref{SMFPT-2}, we observe that $\langle T_R  \rangle$ can also be written as
\bea
\langle T_R  \rangle&=&\frac{\langle \text{min}(T,R) \rangle}{\text{Pr}(T<R)}+\frac{\int_0^\infty~dt~f_R(t)
\int_{\mathcal{D}} d\vec{x}~\tau(\vec{x})~G_0(\vec{x},t)}{\text{Pr}(T<R)} \nonumber \\
&+&\frac{\text{Pr}(R\leq T) \langle W \rangle}{\text{Pr}(T<R)}~.
\label{SLT-T_R}
\eea

\section{Evaluating terms in Eq. (2)}\label{APP_EvalEq2}
\noindent
The expectation value, $\langle \text{min}(T,R) \rangle$, and the probability, $\text{Pr}(T<R)$, in \eref{MFPT-1} are easy to evaluate given the distributions of $T$ and $R$. Indeed, letting $f_T(t)$ and $f_R(t)$ stand for the probability densities of $T$ and $R$ respectively, we see that the cumulative distribution function of $\text{min}(T,R)$
is given by 
\bea
\text{Pr}\left(\text{min}(T,R) \leq t \right)=1-\text{Pr}(T>t)\text{Pr}(R>t),
\label{MinCDF}
\eea
where 
\bea
\text{Pr}(T>t)=\int_t^\infty~dt' f_T(t'),
\eea 
and
\bea
\text{Pr}(R>t)=\int_t^\infty~dt' f_R(t').
\eea
Now, since $\text{min}(T,R)$ is non-negative, the expectation $\langle \text{min}(T,R) \rangle$ can be computed directly from the cumulative distribution function in \eref{MinCDF} as
\bea
\langle \text{min}(T,R) \rangle&=&\int_0^\infty~dt  [1-\text{Pr}\left(\text{min}(T,R)\leq t \right)] \nonumber \\ &=&\int_0^\infty~dt~\text{Pr}(T>t)\text{Pr}(R>t),
\label{mean-min1}
\eea
or, alternatively, using the density of $\text{min}(T,R)$ as
\bea
\langle \text{min}(T,R) \rangle&=&\int_0^\infty~dt~t[f_T(t)\text{Pr}(R>t) \nonumber \\
&+&f_R(t)\text{Pr}(T>t)].
\label{mean-min2}
\eea
Similarly, we see that the probability $\text{Pr}(T<R)$ is given by 
\begin{align}
\text{Pr}(T<R)=\int_0^\infty~dt~f_R(t)\text{Pr}(T<t)=\int_0^\infty~dt~f_T(t)\text{Pr}(R>t)~,
\label{prtleqr}
\end{align}
and note that $\text{Pr}(R \leq T)=1-\text{Pr}(T<R)$.\\

\section{Detailed derivation of  \eref{LT-genericR_main} in the main text}\label{APP_General_Case}
\noindent
We will now derive an exact and general expression for the distribution of the FPT, $T_R$, in Laplace space. 
To this end, we first define two auxiliary random variables
\bea
R_{\text{min}} &\equiv& \{R|R\leq T \}~,\nonumber\\
T_{\text{min}} &\equiv& \{T|T<R \}~.
\label{rmin-tmin-defn}
\eea
In words, $R_{\text{min}}$ is the restart time, $R$, conditioned on the event that restart occurs before the target is found. Similarly, $T_{\text{min}}$ is free-range FPT, $T$, conditioned on the event that the target is found prior to restart. The probability density functions of $R_{\text{min}}$ and $T_{\text{min}}$ are given by \cite{Restart-ND-5,Restart-D-3}
\bea
f_{R_{\text{min}}}(t) &=& \frac{f_R(t) \int_t^\infty~dt'~f_T(t')}{\text{Pr}(R\leq T)}=\frac{f_R(t) \text{Pr}(T>t)}{\text{Pr}(R\leq T)}~,\nonumber\\
f_{T_{\text{min}}}(t) &=& \frac{f_T(t) \int_t^\infty~dt'~f_R(t')}{\text{Pr}(T<R)}=\frac{f_T(t) \text{Pr}(R>t)}{\text{Pr}(T<R)}~.
\label{rmin-tmin-PDF}
\eea
\noindent
To obtain the Laplace transform of $T_R$, we utilize \eref{renewal-1-main} and this gives
\bea
\tilde{T}_R(s)&=&\langle e^{-sT_R}  \rangle  \nonumber \\
&=&
\text{Pr}(T<R)~\left \langle e^{-sT_R}|T<R  \right\rangle+\text{Pr}(R \leq T)~\left \langle e^{-sT_R}|R \leq T  \right\rangle  \nonumber \\
&=& \text{Pr}(T<R)~\left \langle e^{-s \{ T_R|T<R \}} \right \rangle+\text{Pr}(R \leq T)~\left \langle e^{-s\{T_R|R \leq T\}}  \right\rangle .\nonumber \\
\label{SLT1}
\eea
However, by use of \eref{renewal-1-main} and \eref{rmin-tmin-defn} above, we have
\bea
\{ T_R|R \leq T \}&=&\{ R+\tau(\vec{x})+W+T_R'|R \leq T \}\nonumber \\
&=&\{R+\tau(\vec{x})|R \leq T    \}+W+T_R'~,
\eea
and
\bea
\{ T_R|T<R \}=\{ T|T<R \}=T_{\text{min}}~,
\eea
where we have once again utilized the fact that $W$ and  $T_R'$ are independent of $R$ and $T$. Casting these relations back in \eref{SLT1}, we obtain 
\begin{widetext}
\bea
\tilde{T}_R(s)&=& \text{Pr}(T<R)~\left \langle e^{-sT_\text{min}}  \right\rangle+\text{Pr}(R \leq T)~\left \langle e^{-s(W+T_R')-s\{R+\tau(\vec{x})|R \leq T\}}  \right\rangle  \nonumber \\
&=&\text{Pr}(T<R)~\tilde{T}_{\text{min}}(s)+\text{Pr}(R \leq T)~\left\langle e^{-sW}  \right\rangle~\left \langle e^{-sT_R'}  \right\rangle~
\left \langle e^{-s\{R+\tau(\vec{x})|R \leq T\}}  \right\rangle \nonumber \\
&=&\text{Pr}(T<R)~\tilde{T}_{\text{min}}(s)+\text{Pr}(R \leq T)~\tilde{W}(s)~\tilde{T}_R(s)~\left \langle e^{-s\{R+\tau(\vec{x})|R \leq T\}}  \right\rangle~,
\label{STR-Laplace-transform}
\eea
\end{widetext}
where we have again utilized the independence of $W$ and $T_R'$, the fact that $T_R'$ is an IID copy of $T_R$, and further used the shorthand notation $\tilde{Z}(s)$ to denote the Laplace transform of a random variable $Z$. We now observe that
\begin{widetext}
\bea
\left \langle e^{-s\{R+\tau(\vec{x})|R \leq T\}}  \right\rangle  &=& \int_0^{\infty}~dt~f_{R_{\text{min}}}(t)~\left \langle e^{-s\{t+\tau(\vec{x}(t))|T \geq t\}}  \right\rangle \nonumber \\
&=& \int_0^{\infty}~dt~f_{R_{\text{min}}}(t)e^{-st}~\left[ \frac{1}{\text{Pr}(T \geq t)} \int_{\mathcal{D}}~ d\vec{x}~e^{-s\tau(\vec{x})}~G_0(\vec{x},t)  \right]~,
\eea 
\end{widetext}
where we recall that $\text{Pr}(T \geq t)$ is the free-range survival probability. Substituting \eref{rmin-tmin-PDF} into the above we obtain 
\bea
\left \langle e^{-s\{R+\tau(\vec{x})|R \leq T\}}  \right\rangle  = \frac{1}{\text{Pr}(R\leq T)}\int_0^{\infty}~dt~f_{R}(t)e^{-st}~\nonumber \\
\times \int_{\mathcal{D}}~ d\vec{x}~e^{-s\tau(\vec{x})}~G_0(\vec{x},t)~,
\eea 
where we have used $\text{Pr}(T \geq t)=\text{Pr}(T > t)$. Equation (\ref{STR-Laplace-transform}) then reads
\bea
\tilde{T}_R(s)=\text{Pr}(T<R)~\tilde{T}_{\text{min}}(s)+\tilde{W}(s)~\tilde{T}_R(s)~\int_0^{\infty}~dt~f_{R}(t)e^{-st}\nonumber \\
\times \int_{\mathcal{D}} d\vec{x}~e^{-s\tau(\vec{x})}~G_0(\vec{x},t). \nonumber \\
\eea
Rearranging this expression, we obtain an exact and general expression for the FPT, $T_R$, in Laplace space
\bea
\tilde{T}_R(s)=\frac{\text{Pr}(T<R)\tilde{T}_{\text{min}}(s)}{1- \tilde{W}(s)  \int_0^{\infty}~dt~f_{R}(t)e^{-st}~\int_{\mathcal{D}}~ d\vec{x}~e^{-s\tau(\vec{x})}~G_0(\vec{x},t)}~,\nonumber \\
\label{LT-genericR}
\eea
which is \eref{LT-genericR_main} as announced in the main text.

\section{Derivation of \eref{MFPT-2} in the main text}
\label{APP_Eq5}

\noindent
To derive \eref{MFPT-2} in the main text, we simplify \eref{SLT-T_R} by assuming that $f_R(t)=re^{-rt}$, i.e., that restart times are taken from an exponential distribution with
mean $1/r$. First, we use this in \eref{mean-min1} to obtain
\bea
\langle \text{min}(T,R) \rangle
&=&  \int_0^\infty~dt~e^{-rt}\int_t^\infty~dt'~f_T(t') \nonumber \\
&=& \frac{1}{r}-\frac{1}{r} \int_0^\infty dt~ e^{-rt}f_T(t) \nonumber \\
&=&\frac{1-\tilde{T}(r)}{r}~,
\label{mean-min-r}
\eea
where $\tilde{T}(r)$ stands for the Laplace transform of the free-range FPT, $T$, evaluated at $r$.
Similarly, we use \eref{prtleqr} to obtain  
\bea
\text{Pr}(T<R)=\int_0^\infty~dt~f_T(t)e^{-rt}=\tilde{T}(r).
\label{Prob-r}
\eea
Finally, we see that for an exponentially distributed restart time we have
\begin{align}
&\int_0^\infty~dt~f_R(t)
\int_{\mathcal{D}} d\vec{x}~\tau(\vec{x})~G_0(\vec{x},t) \nonumber \\
&=\int_0^\infty~dt~r e^{-rt}
\int_{\mathcal{D}} d\vec{x}~\tau(\vec{x})~G_0(\vec{x},t) \nonumber \\
&= r
\int_{\mathcal{D}} d\vec{x}~\tau(\vec{x})~\int_0^\infty~dt~e^{-rt} G_0(\vec{x},t) \nonumber \\
&= r \int_{\mathcal{D}} d\vec{x}~\tau(\vec{x})~\tilde{G}_0(\vec{x},r),
\label{LTg-r}
\end{align}
where we have defined the Laplace transform of the free-range propagator as
\bea
\tilde{G}_0(\vec{x},r)= \int_0^\infty~ dt~e^{-rt}~G_0(\vec{x},t)~.
\label{G-LT}
\eea
Substituting Eqs. (\ref{mean-min-r}-\ref{LTg-r}) into \eref{SLT-T_R}, we obtain
\bea
\langle T_r  \rangle 
= \frac{1-\tilde{T}(r)}{ \tilde{T}(r)}\frac{1}{r}+\frac{r \int_{\mathcal{D}}~d\vec{x}~\tau(\vec{x})~ \tilde{G}_0(\vec{x},r)}{\tilde{T}(r)}+\frac{1-\tilde{T}(r)}{\tilde{T}(r)}\langle W\rangle~.\nonumber \\
\label{Tr}
\eea Finally, we observe that
\bea
\tilde{T}(r)=\int_0^\infty~dt~e^{-rt}~f_T(t) &=& -\int_0^\infty~dt~e^{-rt}~\frac{d \text{Pr}(T \geq t)}{dt} \nonumber \\
&=&-\int_0^\infty~dt~e^{-rt}~\frac{d}{dt} \int_{\mathcal{D}}~d\vec{x}~G_0(\vec{x},t) \nonumber \\
&=&1-r\int_{\mathcal{D}}~d\vec{x}~\tilde{G}_0(\vec{x},r)~,
\label{tr-propagator}
\eea
where in the last transition we have used integration by parts and the definition in \eref{G-LT}. Multiplying and dividing the second term on the right hand side of \eref{Tr} by $1-\tilde{T}(r)$, using the relation in \eref{tr-propagator}, and setting $\langle \tau(\vec{x})\rangle_r \equiv \int_{\mathcal{D}}~d\vec{x}~\tau(\vec{x})~\phi_r(\vec{x})\equiv\int_{\mathcal{D}}~d\vec{x}~\tau(\vec{x})~\tilde{G}_0(\vec{x},r)/\int_{\mathcal{D}}d\vec{x}~\tilde{G}_0(\vec{x},r)$, we obtain  \eref{MFPT-2} in the main text.


\begin{figure*}[t!]
\includegraphics[width=17.5cm,height=4.5cm]{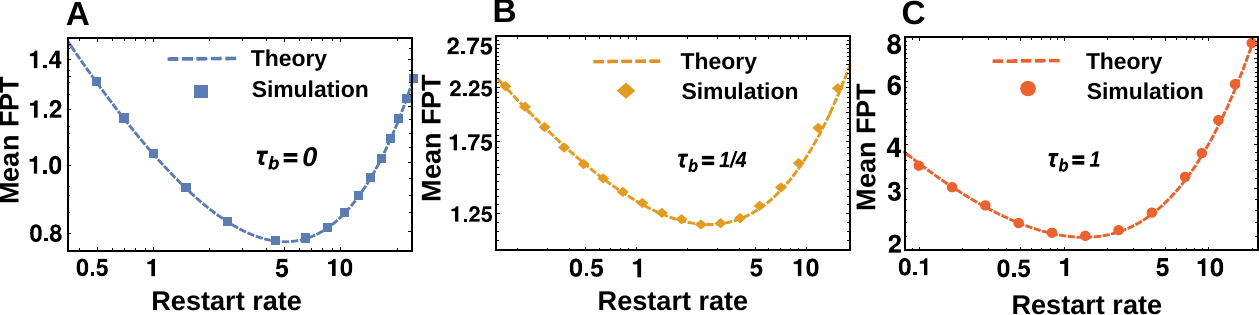}
\caption{The Mean FPT of diffusive search with home returns is plotted vs. the restart rate $r$ for $\tau_d=1/2$ and three different values of $\tau_b$. Dashed lines come from \eref{MFPT-main} in the main text and markers indicate data coming from numerical simulations.}
\label{diffusiontheoryvssimulation}
\end{figure*}


\section{Derivation of \eref{TR-Laplace-transform} in the main text}\label{APP_Eq6}
\noindent
To derive \eref{TR-Laplace-transform} in the main text, we  simplify \eref{LT-genericR} by assuming once again that $f_R(t)=re^{-rt}$, i.e., that restart times are taken from an exponential distribution with mean $1/r$. Using this in \eref{rmin-tmin-PDF}, we immediately find
\bea
\tilde{T}_{\text{min}}(s)=\int_0^\infty dt ~e^{-st}~ f_{T_{\text{min}}}(t)&=&\int_0^\infty dt ~e^{-st}~\frac{f_T(t) \text{Pr}(R>t)}{\text{Pr}(T<R)}\nonumber \\
&=&\frac{\tilde{T}(s+r)}{\tilde{T}(r)}~,
\label{rmintmin}
\eea
where in the last step we have substituted \eref{Prob-r} and used the fact that $\text{Pr}(R>t)=e^{-rt}$. Substituting $f_R(t)=re^{-rt}$, \eref{Prob-r}, and \eref{rmintmin}, into \eref{LT-genericR} and using the definition in \eref{G-LT}, we recover  \eref{TR-Laplace-transform} in the main text
\bea
\tilde{T}_r(s)=\frac{\tilde{T}(s+r)}{1-r~\tilde{W}(s) ~\int_{\mathcal{D}}~d \vec{x}~e^{-s\tau(\vec{x})} \tilde{G}_0(\vec{x},s+r) }~.
\eea


\section{Derivation of \eref{MFPT-main} in the main text}\label{APP_Eq8}
\noindent
To derive \eref{MFPT-main} in the main text, we simplify \eref{MFPT-2} for the case of diffusive home-range search. We first recall that in this case the free-range propagator (starting from the origin and in the presence of a target located at $L$) is given by \cite{FPT1}
\bea
G_0(x,t)=\frac{1}{\sqrt{4\pi D t}} \left( e^{ -\frac{x^2}{4Dt} }-e^{ -\frac{(2L-x)^2}{4Dt} } \right)~.
\label{S-G0-main}
\eea
The Laplace transform of the propagator in \eref{S-G0-main} is given by
\bea
\tilde{G}_0(x,r)&=&\int_0^\infty dt ~e^{-rt}G_0(x,t) \nonumber \\
&=&\frac{1}{\sqrt{4Dr}}\left[ e^{-\sqrt{\frac{r}{D}}|x|}-e^{-\sqrt{\frac{r}{D}}(2L-x)} \right]~.
\eea
Substituting the above expression into \eref{tr-propagator}, we find
\bea
\tilde{T}(r)&=&1-r\int_{\mathcal{D}}~d\vec{x}~\tilde{G}_0(\vec{x},r) \nonumber \\
&=&1-r\int_{-\infty}^L~dx~\tilde{G}_0(x,r) \nonumber \\
&=&e^{-\sqrt{rL^2/D}}=e^{-\sqrt{\tau_d r}}~,
\label{lt-prop-r}
\eea
where we recalled $\tau_d=L^2/D$ from \eref{time-scales}.
Finally, we observe that when the searcher returns home at a constant velocity $v_r$ we have $\tau(x)=|x|/v_r$, and this in turn results in
\bea
r\int_{\mathcal{D}}~d\vec{x}~\tau(\vec{x})~\tilde{G}_0(\vec{x},r)&=&r\int_{-\infty}^L~dx~\frac{|x|}{v_r}~\tilde{G}_0(x,r)\nonumber \\
&=&\tau_b~e^{-\sqrt{\tau_d r}} \left[ \frac{2\sinh(\sqrt{\tau_d r})}{\sqrt{\tau_d r}} -1 \right]~,\nonumber \\
\label{second-term}
\eea
where we have recalled $\tau_b=L/v_r$ from \eref{time-scales} in the main text.
Substituting Eqs. (\ref{lt-prop-r}-\ref{second-term}) into \eref{MFPT-2} and setting $\langle W \rangle=0$, we recover \eref{MFPT-main} in the main text
\bea
\langle T_r  \rangle 
=\frac{1}{r} \left( e^{\sqrt{\tau_d r}}-1 \right) +\tau_b~ \left[ \frac{2\sinh(\sqrt{\tau_d r})}{\sqrt{\tau_d r}} -1 \right]~.
\eea


\section{Corroboration of \eref{MFPT-main} in the main text via numerical simulations}\label{APP_Eq8_cor}
\noindent
In this section, we provide numerical corroboration of \eref{MFPT-main}. In \fref{diffusiontheoryvssimulation}, we  plot the mean FPT for $\tau_d=1/2$ and three different values of $\tau_b$ (indicated on plots) corresponding to those used in \fref{diffusive home-range}B in the main text. In all plots, dashed lines correspond to the exact analytical results coming from \eref{MFPT-main}. These results are corroborated with data coming from numerical simulations (square, diamond, and circle markers). In the simulations, the time step was taken as $\Delta=10^{-5}$ and mean FPTs were estimated based on $10^5$ samples each. As seen from the figure, theory and simulations are in excellent agreement.


\section{Derivation of \eref{optimal-1} in the main text}\label{APP_Eq10}
\noindent
To derive \eref{optimal-1}, we start from \eref{MFPT-main} and set 
\bea
\frac{d}{dr}\langle T_r \rangle =0 ~,
\eea
which gives
\bea
2+2~ \frac{\tau_b}{\tau_d}~ z^{2}\cosh(z)+e^{z}(-2+z)=2 ~\frac{\tau_b}{\tau_d} ~z\sinh (z),
\eea
with $z=\sqrt{r \tau_d}$. Substituting  $2\cosh(z)=e^{z}+e^{-z}$ and $2\sinh(z)=e^{z}-e^{-z}$, we rewrite the above equation as 
\bea
\frac{2}{z^{2}}\frac{1-e^{-z}-\frac{z}{2}}{(1-\frac{1}{z})+(1+\frac{1}{z})e^{-2z}}=\frac{\tau_b}{\tau_d}~.
\label{F_z}
\eea
The left-hand side of \eref{F_z} was defined as $\mathcal{F}(z)$ in the main text.

\section{Expansion of $\mathcal{F}(z)$ around $z=0$}\label{APP_Exp}
\noindent 
We recall the expression for $\mathcal{F}(z)$ from \eref{MFPT-main}
\bea
\mathcal{F}(z)=\frac{2}{z^2}\frac{1-e^{-z}-\frac{z}{2}}{(1-\frac{1}{z})+(1+\frac{1}{z})e^{-2z}}.
\eea
Expanding $\mathcal{F}(z)$ around $z=0$ gives
\bea
\mathcal{F}(z) = \frac{3}{2 z^3}-\frac{2}{5 z}-\frac{1}{8}+O(z)~.
\eea
Thus, in the limit $z\to 0$ we have $\mathcal{F}(z) = \frac{3}{2 z^3} + O(\frac{1}{z})$.

\section{Derivation of \eref{optimalTr*scaling} in the main text}\label{APP_Eq_12}
\noindent
To derive \eref{optimalTr*scaling}, we first write the MFPT from \eref{MFPT-main} at the optimal restart rate
\bea
\langle T_r  \rangle \bigg|_{r=r^*}
=\frac{1}{r^*} \left( e^{\sqrt{\tau_d r^*}}-1 \right) +\tau_b~ \left[ \frac{2\sinh(\sqrt{\tau_d r^*})}{\sqrt{\tau_d r^*}} -1 \right]~.
\label{Tr*}
\eea
To capture the behavior of the MFPT at optimality, at the different limits, we first recall 
\eref{optimalr*scaling} from the main text
\begin{equation}
\begin{array}{l}
r^*/r_0^* \simeq \left\{ \begin{array}{lll}
1 &  & \text{for ~~}\tau_b \ll \tau_d \text{ }\\
 & \text{ \ \ }\\
\left( \frac{3}{2z_0^{*3}} \right)^{2/3}~\left(\frac{\tau_b}{\tau_d}\right)^{-2/3}&  & \text{for~~ }\tau_b \gg \tau_d\text{ ,}
\end{array}\right.\text{ }\end{array}
\label{Soptimalr*scaling}
\end{equation}
where $r_0^*=z_0^{*2}/\tau_d$ and $z^*_0=1.593...$ is the solution of the transcendental equation 
$1-e^{-z}-\frac{z}{2}=0$. In the limit $\tau_b \ll \tau_d$, we have $r^* \simeq r_0^*$, and thus 
\bea
\langle T_{r^{*}}  \rangle \simeq \frac{e^{z_0^*}-1}{z_0^{*2}} \tau_d + \left[ \frac{2\sinh(z_0^*)}{z_0^*}-1 \right] \tau_b \sim \tau_d~.
\label{Tr1*}
\eea
On the other hand, when $\tau_b \gg \tau_d$, we have $r^*/r_0^* \sim \left(\frac{\tau_b}{\tau_d}\right)^{-2/3}$,
so $r^* \sim \tau_d^{-1/3}\tau_b^{-2/3}$, and $\tau_d r^* \sim \left(\frac{\tau_b}{\tau_d}\right)^{-2/3} \ll 1$. Substituting this scaling form into \eref{Tr*}, we find
\bea
\langle T_{r^{*}}  \rangle \simeq \frac{1}{r^*} \left( \sqrt{\tau_d r^*}+\frac{\tau_d r^*}{2} \right)+\tau_b \left( 1+\frac{\tau_d r^*}{3}\right) \sim \tau_b.
\label{Tr2*}
\eea
Equation (\ref{optimalTr*scaling}) then follows immediately from Eqs. (\ref{Tr1*}-\ref{Tr2*}).

\section{Derivation of \eref{criterion-main} in the main text}
\label{derivation13}
\noindent
To derive \eref{criterion-main}, we expand \eref{Tr}
\bea
\langle T_r  \rangle 
= \frac{1-\tilde{T}(r)}{r \tilde{T}(r)}+\frac{r \int_{\mathcal{D}}~d\vec{x}~\tau(\vec{x})~ \tilde{G}_0(\vec{x},r)}{\tilde{T}(r)}+\frac{1-\tilde{T}(r)}{ \tilde{T}(r)}\langle W \rangle, \nonumber \\
\eea
around $r=0$ to obtain
\bea
\langle T_r \rangle =  \langle T \rangle+ \frac{r}{2} \left[  \langle T \rangle^2-\sigma^2(T)  \right]&+ r \int_{\mathcal{D}} ~d\vec{x}~\tau(\vec{x})~\int_0^\infty~dt~G_0(\vec{x},t)\nonumber \\
&+r \langle T \rangle \langle W \rangle+\mathcal{O}(r^2)~,\nonumber \\
\eea
with $\sigma^2(T)=\langle T^2 \rangle-\langle T \rangle^2$ standing for the variance of $T$. Now, the introduction of home returns will decrease the FPT whenever $\langle T_r \rangle<\langle T \rangle$ which is equivalent to 
\bea
\sigma^2(T)-\langle T \rangle^2 >  2\int_{\mathcal{D}} ~d\vec{x}~\tau(\vec{x})~\int_0^\infty~dt~G_0(\vec{x},t)+2\langle T \rangle \langle W \rangle~.\nonumber \\
\eea
Letting $CV=\sigma(T)/\langle T \rangle$ stand for the coefficient of variation, we rearrange the above expression and arrive at the following criterion
\bea
CV^2 > 1+ \frac{2}{\langle T  \rangle^2} \int_{\mathcal{D}}~d\vec{x}~\tau(\vec{x}) \int_0^\infty ~dt~ G_0(\vec{x},t)+\frac{2 \langle W \rangle}{\langle T \rangle}~.\nonumber \\
\label{pre-criterion}
\eea
To get \eref{criterion-main}, we rewrite the second term on the right hand side of \eref{pre-criterion} in a way that resembles the third term in this equation. Observe that 
\bea
&&\frac{1}{\langle T  \rangle} \int_{\mathcal{D}}~d\vec{x}~\tau(\vec{x}) \int_0^\infty ~dt~ G_0(\vec{x},t) \nonumber \\
&=& \int_{\mathcal{D}}~d\vec{x}~\tau(\vec{x}) \left[ \frac{1}{\langle T  \rangle}  \int_0^\infty ~dt~ G_0(\vec{x},t) \right] \nonumber \\
&=& \int_{\mathcal{D}}~d\vec{x}~\tau(\vec{x})~ \phi_0(\vec{x})~,
\eea
where we have again used 
\bea
\phi_0(\vec{x})=\phi_{r=0}(\vec{x})&=&\tilde{G}_0(\vec{x},r=0)/\int_{\mathcal{D}}d\vec{x}~\tilde{G}_0(\vec{x},r=0)\nonumber \\
&=&\frac{1}{\langle T  \rangle}  \int_0^\infty ~dt~ G_0(\vec{x},t)~.
\eea
Once again, we note that  
\bea
\int_{\mathcal{D}}~d\vec{x}~\phi_0(\vec{x}) &=&\int_{\mathcal{D}}~d\vec{x}~
\frac{1}{\langle T  \rangle}  \int_0^\infty ~dt~ G_0(\vec{x},t) \nonumber \\
&=&\frac{1}{\langle T  \rangle} \int_0^\infty ~dt~\int_{\mathcal{D}}~d\vec{x}~G_0(\vec{x},t)\nonumber \\
&=&\frac{1}{\langle T  \rangle} \int_0^\infty ~dt~\text{Pr}(T\geq t)\nonumber
\\ &=& 1,
\eea
which means that $\phi_0(\vec{x})$ is a proper probability density function over the domain $\mathcal{D}$. We thus have 
\bea
\frac{1}{\langle T  \rangle} \int_{\mathcal{D}}~d\vec{x}~\tau(\vec{x}) \int_0^\infty ~dt~ G_0(\vec{x},t) 
= \int_{\mathcal{D}}~d\vec{x}~\tau(\vec{x}) \phi(\vec{x})\nonumber = \langle \tau(\vec{x}) \rangle_0~,
\eea
where the averaging over the return time is done with respect to the probability measure $\phi_0(\vec{x})$. Interpreting $\langle \tau(\vec{x})\rangle_0$ as the mean return time of a home-range searcher in the limit $r\to0$, we substitute the above into \eref{pre-criterion} and recover \eref{criterion-main} in the main text 
\bea
CV^2 > 1+\frac{2 \langle \tau(\vec{x}) \rangle_0}{\langle T  \rangle}+ \frac{2\langle W \rangle }{\langle T \rangle}~.
\label{Scriterion-II}
\eea

\begin{figure*}[t!]
\includegraphics[width=17.5cm,height=4.65cm]{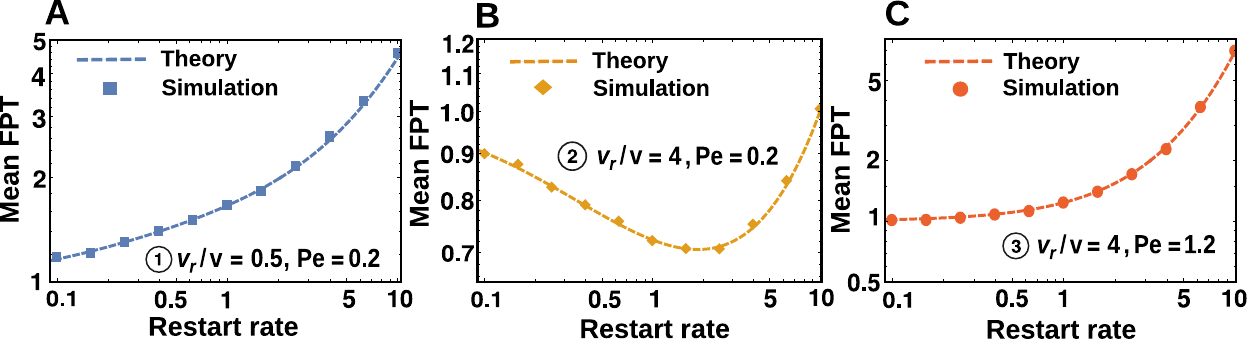}
\caption{The mean FPT of drift-diffusive search with home returns is plotted vs. the restart rate $r$ for the three different sets of parameter that were used in Fig. 4B (indicated on plots). The dashed lines are  exact theoretical results coming from \eref{SMFPT-drift-2} while the markers indicate data coming from numerical simulations.}
\label{driftdiffusiontheoryvssimulation}
\end{figure*}

\section{Drift-diffusive search with home returns}\label{DDS}
\noindent
To derive an expression for the mean FPT of drift-diffusive search with home returns, we simplify \eref{MFPT-2} in the main text for this case. We start from the propagator of the free-range search process
\bea
G_0(x,t)=\frac{1}{\sqrt{4\pi D t}}\left[ e^{-\frac{(x-vt)^2}{4Dt}}-e^{\frac{L v}{D}}e^{-\frac{(x-2L-vt)^2}{4Dt}} \right]~,
\eea
and evaluate its  Laplace transform (at $r$)
\bea
\tilde{G}_0(x,r)&=&\int_0^\infty~dt~e^{-rt}~G_0(x,t)\nonumber \\
&=&\frac{e^{\frac{vx}{2D}}}{\sqrt{v^2+4Dr}} 
\left[ e^{-\frac{|x|}{2D} \sqrt{v^2+4Dr} }-e^{-\frac{|x-2L|}{2D} \sqrt{v^2+4Dr} } \right].\nonumber \\
\eea
Using the above expression, and \eref{tr-propagator}, we compute the Laplace transform of the free-range first-passage time
\bea
\tilde{T}(r)&=&1-r\int_{\mathcal{D}}~d\vec{x}~\tilde{G}_0(\vec{x},r)\nonumber \\
&=&1-r\int_{-\infty}^L~dx~\tilde{G}_0(x,r)\nonumber \\
&=&e^{\frac{Lv}{2D}-\frac{L}{2D}\sqrt{v^2+4Dr}}.
\label{Tr-LT-drift}
\eea
To further proceed, we set $\tau(x)=|x|/v_r$ and compute
\begin{widetext}
\bea
r\int_{\mathcal{D}}~d\vec{x}~\tau(\vec{x})~\tilde{G}_0(\vec{x},r)&=&r\int_{-\infty}^L~dx~\frac{|x|}{v_r}~\tilde{G}_0(x,r) \nonumber \\
&=&\frac{e^{-\frac{L}{D}\sqrt{v^2+4Dr}}}{r v_r\sqrt{v^2+4Dr}} \bigg[ 2Dr \left(-1+e^{\frac{L}{D}\sqrt{v^2+4Dr}}   \right)+v^2 \left( e^{\frac{L}{D}\sqrt{v^2+4Dr}}-1 \right) \nonumber \\
&+& v\sqrt{v^2+4Dr}- \left( v+rL \right) \sqrt{v^2+4Dr} ~e^{\frac{L}{2D}\left(v+\sqrt{v^2+4Dr}\right)} \bigg].
\label{LT-drift2}
\eea
\end{widetext}
Using the above expression and \eref{Tr-LT-drift}, we find
\begin{widetext}
\bea
\frac{r\int_{\mathcal{D}}~d\vec{x}~\tau(\vec{x})~\tilde{G}_0(\vec{x},r)}{\tilde{T}(r)}&=&
\frac{r\int_{-\infty}^L~dx~\frac{|x|}{v_r}~\tilde{G}_0(x,r)}{\exp \bigg[\frac{Lv}{2D}-\frac{L}{2D}\sqrt{v^2+4Dr}\bigg]}
\nonumber \\
&=&\frac{e^{-\frac{Lv}{2D}-\frac{L}{2D}\sqrt{v^2+4Dr}}}{r v_r\sqrt{v^2+4Dr}} \bigg[ 2Dr \left(-1+e^{\frac{L}{D}\sqrt{v^2+4Dr}}   \right)+v^2 \left( e^{\frac{L}{D}\sqrt{v^2+4Dr}}-1 \right) \nonumber \\
&+& \sqrt{v^2+4Dr} \left( v- \left( v+rL \right)  ~e^{\frac{L}{2D}\left(v+\sqrt{v^2+4Dr}\right)}\right) \bigg].
\label{LT-drift3}
\eea
\end{widetext}
Setting $\langle W \rangle=0$ and substituting \eref{Tr-LT-drift} and \eref{LT-drift3} into \eref{MFPT-2}, we find
\bea
\langle T_r \rangle &=& \frac{1}{r} \left[ e^{\frac{L}{2D}\left( \sqrt{v^2+4Dr}-v \right)} -1 \right] \nonumber \\
&+&\frac{e^{-\frac{Lv}{2D}}}{rv_r\sqrt{v^2+4Dr}} \bigg[ (4Dr+2v^2) \sinh \left( \frac{L}{2D}\sqrt{v^2+4Dr} \right) \nonumber \\ 
&+&\sqrt{v^2+4Dr} \left( v e^{-\frac{L}{2D}\sqrt{v^2+4Dr}} -v~e^{\frac{Lv}{2D}}-rL~e^{\frac{Lv}{2D}} \right) \bigg]. \hspace{0.65cm}
\label{SMFPT-drift}
\eea
Recalling $Pe=Lv/2D$, $\tau_d=L^2/D$ and $\tau_b=L/v_r$, we can simplify the above expression further and obtain the following expression for the mean FPT of drift-diffusive search with home returns
\bea
\langle T_r \rangle = \frac{1}{r}\left[ e^{\sqrt{Pe^2+\tau_d r}-Pe}-1 \right] 
+\frac{1}{r} \frac{\tau_b}{\tau_d} ~\mathcal{I}(Pe,\tau_d,r)~,
\label{SMFPT-drift-2}
\eea
where we have introduced the following function
\bea
\mathcal{I}(Pe,\tau_d,r)&=&2 e^{-Pe} \frac{2Pe^2+\tau_d r}{\sqrt{Pe^2+\tau_d r}} \sinh \left[ \sqrt{Pe^2+\tau_d r} \right]  \nonumber \\
&+&2Pe\left[ e^{-\left(Pe+\sqrt{Pe^2+\tau_d r}\right)}-1 \right]-\tau_d r.
\label{SMFPT-drift-3}
\eea
\eref{SMFPT-drift-2} and \eref{SMFPT-drift-3} together constitute a closed form expression for the mean FPT of drift-diffusive search with home returns.

\section{Corroboration of \eref{SMFPT-drift-2} via numerical simulations}\label{DDS_Cor}
In this section, we provide numerical corroboration of \eref{SMFPT-drift-2} for the mean FPT of drift-diffusive search with home returns. Equation (\ref{SMFPT-drift-2}) was used to plot Fig. 4B in the main text. In \fref{driftdiffusiontheoryvssimulation}, we plot the mean FPT from \eref{SMFPT-drift-2} vs. the restart rate for three different sets of parameters (indicated on plots) which correspond to those used in Fig. 4B. In all the plots, dashed lines indicate analytical results coming from \eref{SMFPT-drift-2}. These results are corroborated with data coming from numerical simulations (square, diamond, and circle markers). In the simulations, the time step was taken as $\Delta=10^{-5}$ and mean FPTs were estimated based on $10^5$ samples each. As seen from the figure, theory and simulations are in excellent agreement. 

\section{Derivation of \eref{critical} in the main text}
\noindent
To derive \eref{critical}, we simplify \eref{criterion-main} in the main text for the case of drift-diffusive search with home returns. Starting from the propagator of the free-range search process
\bea
G_0(x,t)=\frac{1}{\sqrt{4\pi D t}}\left[ e^{-\frac{(x-vt)^2}{4Dt}}-e^{\frac{L v}{D}}e^{-\frac{(x-2L-vt)^2}{4Dt}} \right],
\eea
the probability density function of the free-range first-passage time, $T$, can be computed by inverting \eref{Tr-LT-drift} above. One then obtains \cite{FPT1}
\bea
f_T(t)=\frac{L}{\sqrt{4\pi D t^3}} e^{-\frac{(L-vt)^2}{4Dt}}.
\eea
The mean and coefficient of variation of $T$ are then easy to compute. These are given by 
\bea
\langle T \rangle = L/v,
\eea
and
\bea
CV^2=2D/Lv=Pe^{-1},
\eea
where we recalled the definition of the P\'eclet number $Pe=Lv/2D$. With the above at hand, we continue to compute 
\bea
\langle \tau(\vec{x})\rangle_0 = \int_{\mathcal{D}}~d\vec{x}~\tau(\vec{x})~\phi(\vec{x}),
\eea
where $\phi(\vec{x})=\frac{1}{\langle T  \rangle}  \int_0^\infty ~dt~ G_0(\vec{x},t)$. First, we compute 
\bea
\int_0^\infty ~dt~ G_0(x,t)=\frac{1}{v}~ e^{\frac{vx}{2D}} \left[ e^{-\frac{v|x|}{2D}}-e^{-\frac{v|x-2L|}{2D}}  \right], 
\eea
and this allows us to obtain 
\bea
&&\int_{-\infty}^L~dx~\tau(x) ~\int_0^\infty ~dt~ G_0(x,t) \nonumber \\
&=& \int_{-\infty}^L~dx~\frac{|x|}{v_r} ~\int_0^\infty ~dt~ G_0(x,t)  \nonumber \\
&=&
\frac{1}{2v_r v^3} \left[ 4D^2(1-e^{-\frac{Lv}{D}})-2DLv+L^2v^2 \right], 
\eea
and conclude that 
\bea
\langle \tau(\vec{x})\rangle_0 &=& \frac{1}{2Lv_r v^2} \left[ 4D^2(1-e^{-\frac{Lv}{D}})-2DLv+L^2v^2 \right] \nonumber \\
&=&\frac{2D^2}{Lv_r v^2} \left[ 1-Pe+Pe^2-e^{-2Pe} \right].
\eea
Setting $\langle W \rangle=0$ and substituting the expressions for $\langle T \rangle$, $CV^2$, and $\langle \tau(\vec{x})\rangle_0$ into \eref{criterion-main} we obtain
\bea
Pe^{-1}>1+\frac{v}{v_r}\frac{1-Pe+Pe^2-e^{-2Pe}}{Pe^2},
\eea
which can be rearranged to give \eref{critical} in the main text 
\bea
v_r>v_r^*=v \cdot \mathcal{G}(Pe),~~\text{with~~}\mathcal{G}(Pe)
=\frac{1-e^{-2Pe}}{Pe(1-Pe)}-1~.\hspace{0.65cm}
\eea

\end{document}